\documentclass[showkeys,eqsecnum,twocolumn,showpacs,preprintnumbers,amssymb,aps,superscriptaddress]{revtex4}
\usepackage{graphicx}
\usepackage{dcolumn}
\usepackage{bm}
\usepackage{latexsym,epsfig}

\begin{document}

\preprint{\today}

\title{Accurate determination of the electric dipole matrix elements, lifetimes, polarizabilities and light-shift ratios in Ba$^+$ and Ra$^+$}
\vspace{0.5cm}

\author{B. K. Sahoo \footnote{B.K.Sahoo@rug.nl}, R. G. E. Timmermans and K. Jungmann}
\affiliation{KVI, University of Groningen, NL-9747 AA Groningen, The Netherlands}%
\date{\today}
\vskip1.0cm

\begin{abstract}
We have employed the relativistic coupled-cluster theory to calculate the 
electric
dipole matrix elements in the singly ionized barium (Ba$^+$) and radium 
(Ra$^+$). Matrix elements of Ba$^+$ are used to determine the light-shift 
ratios 
for two different wavelengths on which recent experiments are carried out.
By combining the measured light-shift ratios and our calculations, we are
able to estimate possible errors associate with the used matrix elements
in Ba$^+$. Static polarizabilities of the low-lying states and lifetimes
of the first excited P-states are also determined using these matrix 
elements in Ba$^+$. A similar approach has been followed to estimate the 
lifetimes and light-shift ratios in Ra$^+$.
\end{abstract}


\maketitle

\section{Introduction}
Recently, both the singly ionized barium (Ba$^+$) and (Ra$^+$) are proposed
as suitable candidates for the parity non-conservation (PNC) experiments 
\cite{fortson1,wansbeek1}. These candidates are also suitable for the 
atomic clock experiments \cite{fortson2,bijaya1,bijaya2}. Accurate determination of the electric dipole (E1) matrix elements are essential in achieving sub-one
percent PNC amplitudes in the above candidates \cite{bijaya3,wansbeek1}.
They are also crucial in determining the dipole polarizabilities accurately
which are required to estimate shifts due to the stray electric fields applied 
\cite{bijaya2,itano} during the measurements in the above proposed experiments.
There is no direct method to measure them experimentally. Mainly, these
results are estimated from the branching ratio (lifetime)
measurements from various states. However, when branching ratios of two or more
different channels from a given state are relatively significant, it is not
possible to estimate the E1 matrix elements precisely. There are a few 
branching ratio \cite{kastberg,reader,gallagher1} and lifetime \cite{kuske,pinnington,andrae} measurements of the first P-states available in Ba$^+$, 
although they are not very precise to
estimate the E1 matrix elements accurately. On the otherhand, such experiments
are not available yet in Ra$^+$.

It is also possible to calculate the above matrix elements accurately using 
a highly potential many-body method. Ba$^+$ and Ra$^+$ are heavy systems
implying both the electron correlation and relativistic effects will be much
stronger in these systems. Variety of relativistic many-body methods have 
already
been employed to calculate the E1 matrix elements in the considered systems 
\cite{dzuba,geetha,safronova1,johnson,safronova2}. But the results obtained
by all these works differ at significant precision for which they can be used
to determine sub-one percent accuracy of the PNC amplitudes or other experimentally measured quantities
like light-shift ratios \cite{fortson3,fortson4} in the considered systems.

In this work, we have employed the relativistic coupled-cluster (RCC) method
to account more electron correlation effects through the non-linear terms
for the determination of many E1 matrix elements in both Ba$^+$ and Ra$^+$. These results
are then used to evaluate the light-shift ratios in Ba$^+$ at two wavelengths 
on which experiments were conducted \cite{fortson3,fortson4}. By analyzing the
trends of contributions to the light-shift ratios and transition probabilities,
we try to give the upper limits to some of the important matrix elements by 
combining with the experimental result in that system.
The matrix elements are further used to determine the lifetimes and static 
dipole polarizabilities in Ba$^+$. By following the same procedure, we 
determine in Ra$^+$ the lifetimes of the first excited P-states and light-shift
ratios at the same wavelengths in which experiments were carried out in Ba$^+$.

\section{Theory and method of calculations}
The shift of energy in an atomic state $|\gamma, J, M \rangle$, where 
$J$ represents the total angular momentum of the state with its azimuthal 
component $M$ and $\gamma$ is the additional index representing other 
required quantum numbers, due to non-resonant ac light in an average 
period of light oscillations and neglecting the mixing of the magnetic
sub-levels is given by \cite{budker,stalnaker}
\begin{eqnarray}
\Delta {\cal E} (\gamma, J, M) &=& - \frac{\alpha_0}{2} |{\bf \vec E}|^2 - i \frac{\alpha_1}{2} \frac{M}{J} (i |{\bf \vec E}^* \times {\bf \vec E}|) - \nonumber \\
 && \frac{\alpha_2}{2} \left ( \frac{3M^2-J(J+1)}{J(2J-1)} \right ) \frac
{3 E_z^2 - |{\bf \vec E}|^2}{2}, \ \ \ \ \
\label{eqn1}
\end{eqnarray}
where ${\bf \vec E}$ is the applied vector electric field and $E_z$ is its 
magnitude in the z-component. Here $\alpha_0$, $\alpha_1$ and $\alpha_2$ are
the scalar, vector and tensor polarizabilities of the state $|\gamma, J, M \rangle$.
This expression differs by the term containing vector polarizability when the
dc electric field is applied. 

Using the first-order time-dependent perturbation theory and assuming large 
detuning from resonance with the applied electric field, the scalar, vector
and tensor dynamic polarizabilities of $|\gamma, J, M \rangle$ given as \cite{stalnaker,bonin}
\begin{eqnarray}
\alpha_0(\gamma, J, M) &=& - \frac{2}{3[J]} \sum_{K \ne J} \frac{E_J -E_K}{(E_J -E_K)^2 - \omega^2} \nonumber \\ && |\langle \gamma' K ||D|| \gamma J \rangle |^2 \ \ \phi_0(J,K) \label{eqn21} \\
\alpha_1(\gamma, J, M) &=& - \frac{1}{[J]} \sum_{K \ne J} \frac{\omega}{(E_J -E_K)^2 - \omega^2} \nonumber \\ && |\langle \gamma' K ||D|| \gamma J \rangle |^2 \ \ \phi_1(J,K) \label{eqn22} \\
\alpha_2(\gamma, J, M) &=& - \frac{2}{3[J]} \sum_{K \ne J} \frac{E_J -E_K}{(E_J -E_K)^2 - \omega^2} \nonumber \\ && |\langle \gamma' K ||D|| \gamma J \rangle |^2 \ \ \phi_2(J,K),
\label{eqn23}
\end{eqnarray}
respectively, where $D$ is the E1 operator, $[J]$ is the degeneracy factor equal to $2J+1$, 
$ \gamma'$ and $K$ represent different states than $|\gamma, J, M \rangle$ 
with opposite parity, matrix elements with double bars represent the reduced 
matrices, $E_{J/K}$ represent energies of the corresponding states in the 
absence of the external electric field and $\omega$ is the frequency of the 
applied electric field. 

The scalar factors $\phi_i(J,K)$'s ($i=0,1,2$) multiplied in the above 
equations are defined as
\begin{eqnarray}
\phi_0(J,K) &=& \delta_{J-1,K} + \delta_{J,K} + \delta_{J+1,K} \\
\phi_1(J,K) &=& - \frac{1}{J} \delta_{J-1,K} - \frac{1}{J(J+1)} \delta_{J,K} \nonumber \\ && + \frac{1}{J+1} \delta_{J+1,K} \\
\phi_2(J,K) &=& - \delta_{J-1,K} + \frac{2J-1}{J+1} \delta_{J,K} \nonumber \\ && - \frac{J(2J-1)}{(J+1)(2J+3)} \delta_{J+1,K}.
\label{eqn3}
\end{eqnarray}
The static dipole polarizabilities can be determined by substituting $\omega$ as zero in the above equations.

In fact, the vector shift is maximal than other two shifts for the
purely circular polarized light when it is aligned with any existing magnetic
field \cite{mathur,tannoudji}. In this case, it is possible to measure the
ratio of light shifts instead of measuring the light shifts or light intensities
directly. Mathematically, it is equivalent to say as the following; the ratio
of the shift in energies of two different states ($|\gamma_i, J_i, M_i \rangle$ and $|\gamma_f, J_f, M_f \rangle$) is given by
\begin{eqnarray}
R &=& \frac{\Delta {\cal E} (\gamma_i, J_i, M_i)}{\Delta {\cal E} (\gamma_f, J_f, M_f)} \nonumber \\
 &=& \frac {\alpha_1(\gamma_i, J_i, M_i)}{\alpha_1(\gamma_f, J_f, M_f)},
\label{eqn4}
\end{eqnarray}
which is known as the light-shift ratio in the literature \cite{fortson3,fortson4}.

The probability coefficient (in $s^{-1}$) due to the allowed transition is given by
\begin{eqnarray}
A^{\text{E1}}_{f \rightarrow i} &=& \frac {2.02613 \times 10^{18} }{\lambda^3 [J_f]} S_{f \rightarrow i}^{\text{E1}},
\label{eqn5}
\end{eqnarray}
where $\lambda$ (\AA) and $S_{f \rightarrow i}^{\text{E1}} (= |\langle f || D || i \rangle|^2)$ (au) are the wavelengths and strengths of the corresponding 
transitions, respectively.

\begin{figure}[h]
\includegraphics[width=8.5cm,clip=true]{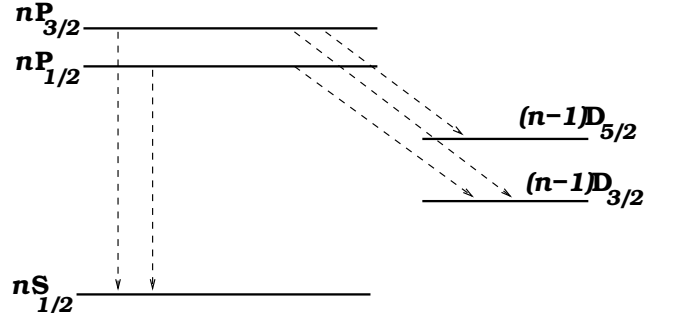}
\caption{Schematic low-lying energy level diagrams and decay of the P-states in Ba$^+$ and Ra$^+$. $n=6$ and $n=7$ in Ba$^+$ and Ra$^+$, respectively.}
\label{fig1}
\end{figure}
As seen in Fig. \ref{fig1}, the first excited P-states in both Ba$^+$ and Ra$^+$
can decay either to the ground S-states directly or via the excited 
D-states through the allowed transitions. The lifetimes of these P-states
can be evaluated from the inverse of the corresponding net transition 
probabilities due to all possible transition channels. Therefore, the net
transition probabilities for the P-states are given by
\begin{eqnarray}
A_{n\text{P}_{1/2}}&=& A_{n\text{P}_{1/2}\rightarrow n\text{S}_{1/2}} + A_{n\text{P}_{1/2}\rightarrow (n-1)\text{D}_{3/2}} \\
A_{n\text{P}_{3/2}}&=& A_{n\text{P}_{3/2}\rightarrow n\text{S}_{1/2}} + A_{n\text{P}_{3/2}\rightarrow (n-1)\text{D}_{3/2}} \nonumber \\ && + A_{n\text{P}_{3/2}\rightarrow (n-1)\text{D}_{5/2}}
\label{eqn6}
\end{eqnarray}
and hence, the lifetimes (in $s$) of these states are given by
\begin{eqnarray}
\tau_{n\text{P}_{1/2}}&=& \frac{1}{A_{n\text{P}_{1/2}}} \\ 
\tau_{n\text{P}_{3/2}}&=& \frac{1}{A_{n\text{P}_{3/2}}},
\label{eqn7}
\end{eqnarray}
where $n=6$ and $n=7$ in Ba$^+$ and Ra$^+$, respectively.

Since the transition wavelengths can be determined from the excitation energies
from the corresponding transitions, it is clear from the above equations 
that both the polarizabilities and transition probabilities depend upon the
excitation energies (or wavelengths) and E1 matrix elements. Since our 
motivation in the present work is to verify the accuracies of the E1 matrix
elements, we consider the experimental (observational) excitation energies
(wavelengths) for our purpose and analyze only the role of the E1 matrix 
elements from our calculations.

We employ the RCC theory for the Dirac-Coulomb Hamiltonian in the single and 
double excitations approximation along with the important triple excitations
(CCSD(T) method) to calculate the atomic wave functions in both Ba$^+$ and 
Ra$^+$. In this procedure, we express the one valence ($v$) configuration
state function as \cite{debasish,lindgren}
\begin{eqnarray}
|\Psi_v \rangle = e^T \{ 1 + S_v \} |\Phi_v\rangle,
\label{eqn8}
\end{eqnarray}
where the reference state $|\Phi_v\rangle$ is constructed by appending the 
corresponding valence orbital to the closed-shell Dirac-Fock (DF) wave 
function $|\Phi_0\rangle$; which in second quantization formalism is expressed
as $|\Phi_v\rangle=a_v^{\dagger}|\Phi_0\rangle$. In the above equation, $T$
and $S_v$ are the core and valence-core correlation operators. In the linear
approximation, the above equation yields the form
\begin{eqnarray}
|\Psi_v \rangle = \{ 1 + T + S_v \} |\Phi_v\rangle.
\label{eqn9}
\end{eqnarray}
As shown in the earlier works, contributions (electron correlation effects)
from higher triple and quadrupole 
excitations arise through the non-linear terms in the CCSD(T) method and their
contributions to the various properties in the heavy systems are significantly
large \cite{bijaya4,bijaya5}.

The reduced transition matrix elements $\langle \Psi_f ||D||\Psi_i \rangle$ 
between two states $|\Psi_f \rangle$ and $|\Psi_i \rangle$ are evaluated by
\begin{widetext}
\begin{eqnarray}
\langle \Psi_f ||D||\Psi_i \rangle = \frac {\langle \Phi_f || \{ 1 + S_f^{\dagger} \} \overline{D} \{ 1 + S_i \} ||\Phi_i\rangle}{\sqrt{\langle \Phi_f || \{ 1 + S_f^{\dagger} \} \overline{N} \{ 1 + S_f \} ||\Phi_f\rangle \langle \Phi_i || \{ 1 + S_i^{\dagger} \} \overline{N} \{ 1 + S_i \} ||\Phi_i\rangle}} 
\label{eqn10}
\end{eqnarray}
\end{widetext}
where $\overline{D}=e^{T^{\dagger}} D e^T$ and $\overline{N}=e^{T^{\dagger}} e^T$ are the non-truncative series in the ordinary RCC theory. However, in the CCSD(T) method, only up to the effective three-body terms, when they are expanded
using the general Wick's theorem, of these series will contribute to the above 
expression. These terms are calculated step by step and sandwiched between 
the necessary $S_v$ operators to evaluate the matrix elements. In order to
estimate the errors associate with the calculated E1 matrix elements, we check
the convergence of these results using both the length and velocity gauge 
expressions which at the single orbital level are given by
\begin{widetext}
\begin{eqnarray}
\langle f ||d||i \rangle = \langle \kappa_f || C^{(1)} || \kappa_i \rangle 
\int_0^{\infty} dr \left \{ r [P_f(r) P_i(r) + Q_f(r) Q_i(r)] 
 - \frac{\alpha}{5} (\epsilon_i- \epsilon_f) \left ( \frac{(\kappa_i-\kappa_f)}{2}[P_f(r) Q_i(r) + Q_f(r) P_i(r)] + [P_f(r) Q_i(r) - Q_f(r) P_i(r)] \right ) \right \}
\label{eqn11}
\end{eqnarray}
in the length form and
\begin{eqnarray}
\langle f ||d||i \rangle = \langle \kappa_f || C^{(1)} || \kappa_i \rangle 
\frac{1}{\alpha (\epsilon_i- \epsilon_f)} \int_0^{\infty} dr \left \{ (\kappa_i-\kappa_f) [P_f(r) Q_i(r) + Q_f(r) P_i(r)] 
 - [P_f(r) Q_i(r) - Q_f(r) P_i(r)] \right \},
\label{eqn12}
\end{eqnarray}
\end{widetext}
in the velocity form, respectively. In these expressions, $C^{(1)}$ is the 
Racah vector with rank one, $P_{i/f}(r)$ and $Q_{i/f}(r)$ are the large
and small components of the Dirac single particle orbitals, $\kappa_{i/f}$ 
are the relativistic quantum numbers, $\epsilon_{i/f}$ are the single particle
energies and $\alpha$ is the fine structure constant.

\section{Results and discussions}
\subsection{Ba$^+$}
We present our calculated E1 matrix elements of Ba$^+$ in Table \ref{tab1}. 
The reported error bars in these results except from the matrix elements
related to the D- and F- states are estimated from the differences between the 
length and velocity gauge results. It was rather difficult to converge the 
velocity 
gauge results for the D- and F- states, therefore we consider the difference
between the converged length gauge results using the CCSD(T) method and just
by single and double excitations approximation (CCSD method), which are assumed
as the upper limit to the contributions due to the neglected excitations,
as the error bars. However, these error bars are just used to estimate errors 
in the light-shift ratio calculations in the considered system and we analyze
the accuracy of these results later by fitting with the experimentally 
measured values of the light-shift ratios at two different wavelengths as 
discussed below.
\begin{table}[h]
\caption{Absolute magnitudes of the reduced dipole matrix elements in Ba$^{+}$. Estimated error bars from this work given inside the parenthesis.}
\begin{ruledtabular}
\begin{center}
\begin{tabular}{lccccc}
Transition & Present & Ref. \cite{dzuba} & Ref. \cite{geetha} & Ref. \cite{safronova1} & Ref. \cite{johnson} \\
\hline \\
  &   &  &  &  &  \\
$6p_{1/2}\rightarrow 6s_{1/2}$ & 3.36(1) & 3.310 & 3.3266 & 3.3357 & 3.300 \\ 
$7p_{1/2}\rightarrow 6s_{1/2}$ & 0.10(1) & 0.099 & 0.1193 & 0.0621 &    \\ 
$8p_{1/2}\rightarrow 6s_{1/2}$ & 0.11(5) & 0.115 & 0.4696 &     &    \\ 
$6p_{3/2}\rightarrow 6s_{1/2}$ & 4.73(3) & 4.674 & 4.6982 & 4.7065 & 4.658 \\ 
$7p_{3/2}\rightarrow 6s_{1/2}$ & 0.17(5) & 0.035 & 0.3610 & 0.0868 &    \\ 
$8p_{3/2}\rightarrow 6s_{1/2}$ & 0.11(5) & 0.073 & 0.5710 &     &    \\ 
$6p_{1/2}\rightarrow 7s_{1/2}$ & 2.44(4) & 2.493 & 2.3220 &     &    \\ 
$6p_{1/2}\rightarrow 8s_{1/2}$ & 0.66(5)& 0.705 & 0.7283 &     &    \\ 
$6p_{3/2}\rightarrow 7s_{1/2}$ & 3.80(2) & 3.882 & 3.6482 &     &    \\ 
$6p_{3/2}\rightarrow 8s_{1/2}$ & 0.97(5) & 1.025 & 1.0518 &     &    \\ 
$6p_{1/2}\rightarrow 5d_{3/2}$ & 3.11(3) & 3.055 & 2.9449 &  & 3.009 \\ 
$7p_{1/2}\rightarrow 5d_{3/2}$ & 0.28(2) & 0.261 & 0.3050 &     &    \\ 
$8p_{1/2}\rightarrow 5d_{3/2}$ & 0.13(2) & 0.119 & 0.1121 &     &    \\ 
$6p_{3/2}\rightarrow 5d_{3/2}$ & 1.34(2) & 1.334 & 1.2836 &     & 1.312 \\ 
$7p_{3/2}\rightarrow 5d_{3/2}$ & 0.16(1) & 1.472 & 0.1645 &     &    \\ 
$8p_{3/2}\rightarrow 5d_{3/2}$ & 0.07(2) & 0.070 & 0.0650 &     &    \\ 
$4f_{5/2}\rightarrow 5d_{3/2}$ & 3.75(11) &    &     &     &   \\ 
$5f_{5/2}\rightarrow 5d_{3/2}$ & 1.59(8) &    &     &     &   \\ 
$6f_{5/2}\rightarrow 5d_{3/2}$ & 0.17(2) &    &     &     &   \\ 
$6p_{3/2}\rightarrow 5d_{5/2}$ & 4.02(7) & 4.118 & 3.9876 &     & 4.057 \\ 
$7p_{3/2}\rightarrow 5d_{5/2}$ & 0.46(1) & 0.432   & 0.4788 &   &    \\ 
$8p_{3/2}\rightarrow 5d_{5/2}$ & 0.21(2) & 0.206  & 0.1926  &   &    \\ 
$4f_{5/2}\rightarrow 5d_{5/2}$ & 1.08(4) &    &     &     &   \\ 
$5f_{5/2}\rightarrow 5d_{5/2}$ & 0.45(7) &    &     &     &   \\ 
$6f_{5/2}\rightarrow 5d_{5/2}$ & 0.15(2) &    &     &     &   \\ 
$4f_{7/2}\rightarrow 5d_{5/2}$ & 4.84(5) &    &     &     &   \\ 
$5f_{7/2}\rightarrow 5d_{5/2}$ & 2.47(6) &    &     &     &   \\ 
$6f_{7/2}\rightarrow 5d_{5/2}$ & 1.04(7) &    &     &     &   \\ 
$6p_{1/2}\rightarrow 6d_{3/2}$ & 4.89(10) &  &  &  &  \\ 
$6p_{1/2}\rightarrow 7d_{3/2}$ & 1.50(8) &  &  &     &  \\ 
$6p_{3/2}\rightarrow 6d_{3/2}$ & 2.33(7) &  &  &  &  \\ 
$6p_{3/2}\rightarrow 7d_{3/2}$ & 0.67(4) &  &  &  &  \\ 
$6p_{3/2}\rightarrow 6d_{5/2}$ & 6.91(21) &  &  &     &  \\ 
$6p_{3/2}\rightarrow 7d_{5/2}$ & 2.01(5) &  &  &     &  \\ 
\end{tabular}
\end{center}
\end{ruledtabular}
\label{tab1}
\end{table}

\begin{table*}[h]
\caption{Dynamic vector polarizabilities (in au) at two different wavelengths in Ba$^{+}$.}
\begin{ruledtabular}
\begin{center}
\begin{tabular}{lcccc}
Intermediate & \multicolumn{2}{c}{$6s_{1/2}$} & \multicolumn{2}{c}{$5d_{3/2}$}\\
States ($K$)  &  $\lambda=514.53$nm  &  $\lambda=1111.68$nm &  $\lambda=514.53$nm  &  $\lambda=1111.68$nm \\
\hline \\
   &  &  &  \\
$6p_{1/2}$ & $-$978.638 & $-$45.080 &  48.786 & $-$20.419 \\
$7p_{1/2}$ & $-$0.014 & $-$0.006 & $-$0.035  & $-$0.014   \\
$8p_{1/2}$ & $-$0.010 & $-$0.004 & $-$0.004 & $-$0.002   \\
$6p_{3/2}$ & 305.338 & 36.717 & 6.112 & $-$1.121 \\
$7p_{3/2}$ & 0.019 & 0.008 & $-$0.004 & $-$0.002   \\
$8p_{3/2}$ & 0.005 & 0.002 & $-5\times10^{-4}$ & $-4\times10^{-4}$   \\
$4f_{5/2}$ &  &    & 3.987 & 1.541 \\
$5f_{5/2}$ &  &    & 0.153 &  0.162   \\
$6f_{5/2}$ &  &    & 0.004 & 0.002   \\
Others    & 0.301 & 0.137 & 0.243 & 0.177   \\
\hline \\
Total & $-$672.994(1.944)& $-$8.223(201) & 59.298(1.375) & $-$19.675(323)\\
\end{tabular}
\end{center}
\end{ruledtabular}
\label{tab2}
\end{table*}
We also compare our results with the earlier works in Table \ref{tab1}. As seen,
most of the important low-lying E1 matrix elements from various works differ 
at the second decimal places which means the difference is mainly due to 
the different many-body methods employed and they may not be due to the 
numerical methods used in the calculations. That is the
reason why we have presented our results up to the second decimal places in the 
above table. As a matter of fact, correct results up to the second decimal
places are crucial in obtaining the light-shift ratios correctly as will
be shown below. First, we would like to briefly discuss here the important
differences in the various works presented in Table \ref{tab1}. Geetha et al.
\cite{geetha} have employed the same approach like ours, but they had used 
part numerical orbitals using GRASP and part analytical orbitals using 
Gaussian type orbitals (GTOs). In the present work, we have considered pure 
analytical orbitals using GTOs. Another difference between these two works is 
that $\overline{D}$ of Eq. (\ref{eqn10}) is truncated only at the effective 
one-body 
terms by Geetha et al. \cite{das}, where other higher order terms have been 
considered in the present calculation. The main difference in the methods 
applied by Dzuba et al. \cite{dzuba} and ours have been discussed by Geetha
et al. in their work \cite{geetha}. In summary, they have employed the Green's
function technique which is also an all order
perturbative method like ours. In their work, the orbitals are 
also obtained analytically like the present work. However, the level of
approximation in both the works could be different which cannot be obviously
distinguished. Iskrenova-Tchoukova and Safronova \cite{safronova1} have used linearized CCSD 
method with important triple excitations and orbitals have been constructed 
using the B-spline basis. As has been pointed out earlier, the non-linear terms
can incorporate higher order correlation effects which are necessary \cite{bijaya4,bijaya5} to account 
in the considered heavy system. Again, the procedure to consider partial 
triple excitations in their method and in the present work is different. We
consider effects of partial triple excitations by estimating each time their 
contributions to the energy of the corresponding valence state and then
solve the CCSD method amplitudes self-consistently which involves the above
energy. However, Iskrenova-Tchoukova and Safronova include them directly in the amplitude
determining equations. Guet and Johnson have just applied the relativistic 
many-body method at the second order approximation (MBPT(2) method) using 
the B-spline basis to calculate their results in contrast to our method which 
is all order in nature \cite{lindgren,bijaya4}.

We have used our calculated reduced E1 matrix elements and experimental energies
 \cite{moore} in Eq. (\ref{eqn22}) to obtain the dynamic vector dipole 
polarizabilities ($\alpha_1$) of the $6\text{S}_{1/2}$ and $5\text{D}_{3/2}$ states in 
Ba$^+$ at two different wavelengths (of external electric fields) and they 
are presented in Table \ref{tab2}. Only $6\text{P}_{1/2}$, $7\text{P}_{1/2}$, $8\text{P}_{1/2}$, $6\text{P}_{3/2}$, $7\text{P}_{3/2}$, $8\text{P}_{3/2}$, $4\text{F}_{5/2}$, $5\text{F}_{5/2}$ and $6\text{F}_{5/2}$ intermediate states ($K$) 
have been taken into account as they seem to be the most important contributing
states as seen from this table. It can be noticed that the magnitude of 
$\alpha_1$s of the $6\text{S}_{1/2}$ state are larger than the $5\text{D}_{3/2}$ state at both the wavelengths, but their differences are smaller when
the frequency of the external field is small. Again, the sign of these
quantities are different at the smaller wavelength (514.53 nm), whereas they 
are opposite at the large wavelength (1111.68 nm). For both the states, the 
largest contribution comes from the $6\text{P}_{1/2}$ state. The next highest
contribution arises from the $6\text{P}_{3/2}$ state. The third largest
contributions to $\alpha_1$ of the $5\text{D}_{3/2}$ state comes from the 
$4\text{F}_{5/2}$ state. Surprisingly, the rest of the contributions are 
dominated 
by the higher F-states compared to the P-states. We have given contributions
from pure core correlations, core-valence correlations and higher states 
those are neglected as "Others". The core correlation effects are around
0.319 au and 0.147 au at the wavelengths 514.53 nm and 1111.68 nm, 
respectively. These contributions are evaluated using the MBPT(2) method.
The possible errors due to the omission of the higher states may come from the
neglected F-states which seems to be important in the determination of
$\alpha_1$ of the $5\text{D}_{3/2}$ state. Therefore, an approach similar
to the one applied to determine the dynamic scalar and tensor polarizabilities
\cite{wansbeek2} will be more appropriate to use for the high precision 
light-shift ratio studies in the present system.

Now considering the above results of  $\alpha_1$ for both the $6\text{S}_{1/2}$
and $5\text{D}_{3/2}$ states and using Eq. (\ref{eqn4}), we obtain the 
light-shift ratios in Ba$^+$ as $R^{\lambda=514.53nm} = -11.36(20)$ and
$R^{\lambda=1111.68nm} = 0.418(3)$, where core correlations contribute around
0.6\% and 1\% at the wavelengths 514.53 nm and 1111.68 nm, respectively. 
The corresponding experimental results are $R^{\lambda=514.53nm} = -11.494(13)$ 
and $R^{\lambda=1111.68nm} = 0.4176(8)$ at the wavelengths 514.53 nm and 1111.68 nm, respectively \cite{fortson3,fortson4}. Our results match well with the
corresponding experimental results. It shows that accounting core correlation 
effects are essential for the high precision results in the light-shift studies,
although their magnitudes seem to be very small in comparison to the $\alpha_1$ of the $6\text{S}_{1/2}$ state. Especially, their contributions are relatively
larger at the higher wavelength (smaller frequency). The errors quoted in our
light-shift ratio calculations are obtained from the error bars given for the
E1 matrix elements in Table \ref{tab1}.

We are now in a position to analyze the accuracy of the E1 matrix elements
by studying their contributions to the light-shift ratio calculations. Since 
contributions from the $6\text{P}_{1/2}$ state seems larger, we first analyze matrix elements involving this state. From 
Table \ref{tab1}, we can find that different theoretical predictions of the 
magnitude (sign is irrelevant for our observable) $\langle 6p_{1/2} || D || 6s \rangle$ range from 3.30 au to 3.37 au (along with the error bar). Now keeping 
all
other matrix elements unchanged from our calculations, we observe that light-shift ratios
vary from $R^{\lambda=514.53nm} = -10.77$ and $R^{\lambda=1111.68nm} = 0.337$
to $R^{\lambda=514.53nm} = -11.45$ and $R^{\lambda=1111.68nm} = 0.432$. With
$\langle 6p_{1/2} || D || 6s\rangle=3.33$, which is reported by Geetha
et al. \cite{geetha} and Iskrenova-Tchoukova and Safronova \cite{safronova1}, we observe
$R^{\lambda=514.53nm} = -11.06$ and $R^{\lambda=1111.68nm} = 0.337$. 
Keeping the value of the above matrix element constant, we consider now the 
range of $\langle 6p_{3/2} || D || 6s \rangle$ as 4.66 au to 4.76 au and
observe from $R^{\lambda=514.53nm} = -11.21$ and $R^{\lambda=1111.68nm} = 0.432$ to $R^{\lambda=514.53nm} = -10.99$ and $R^{\lambda=1111.68nm} = 0.353$. To improve these results, we fix now $\langle 6p_{3/2} || D || 6s \rangle=4.70$ and
vary $\langle 6p_{1/2} || D || 6s\rangle$ result again. We find that 
when we consider $\langle 6p_{1/2} || D || 6s\rangle=3.35$ and $\langle 6p_{1/2} || D || 6s\rangle=3.37$ for $\langle 6p_{3/2} || D || 6s \rangle=4.70$, we get $R^{\lambda=514.53nm} = -11.32$ and $R^{\lambda=1111.68nm} = 0.428$ and $R^{\lambda=514.53nm} = -11.51$ and $R^{\lambda=1111.68nm} = 0.455$, respectively. Since the former result, is little close with the experimental results for both
the wavelengths, we now keep $\langle 6p_{1/2} || D || 6s\rangle=3.35$ and 
increase $\langle 6p_{3/2} || D || 6s \rangle$ result and get $R^{\lambda=514.53nm} = -11.27$ and $R^{\lambda=1111.68nm} = 0.412$ and $R^{\lambda=514.53nm} = -11.22$ and $R^{\lambda=1111.68nm} = 0.396$ for 4.72 au and 4.74 au, respectively.
Therefore, we conclude now that $\langle 6p_{3/2} || D || 6s \rangle$ will be around
4.73 au. Now keeping $\langle 6p_{1/2} || D || 6s\rangle=3.35$ and $\langle 6p_{3/2} || D || 6s\rangle=4.73$ when we vary $\langle 6p_{1/2} || D || 5d_{3/2}\rangle$ from 3.00 au to 3.15 au, we get from $R^{\lambda=514.53nm} = -11.93$ and $R^{\lambda=1111.68nm} = 0.435$ to $R^{\lambda=514.53nm} = -11.02$ and $R^{\lambda=1111.68nm} = 0.394$. Results close to the experimental measurements are able to produce only when $\langle 6p_{1/2} || D || 5d_{3/2}\rangle$ is around 3.08 au.
Now we keep $\langle 6p_{1/2} || D || 6s\rangle=3.35$, $\langle 6p_{3/2} || D || 6s\rangle=4.73$ and $\langle 6p_{1/2} || D || 5d_{3/2}\rangle=3.08$ then
vary $\langle 6p_{3/2} || D || 5d_{3/2}\rangle$ matrix element from 
1.28 au to 1.36 au. We get light-shift ratios close to experimental results 
when $\langle 6p_{3/2} || D || 5d_{3/2}\rangle$ is around 1.32 au. However, 
light-shift results vary slowly in the given range of this matrix element while
the previous matrix elements considered earlier play the crucial roles in 
deciding the final results.
We now vary $\langle 4f_{5/2} || D || 5d_{3/2}\rangle$ matrix element
from 3.50 au to 4.00 au by considering $\langle 6p_{1/2} || D || 6s\rangle=3.35$, $\langle 6p_{3/2} || D || 6s\rangle=4.73$, $\langle 6p_{1/2} || D || 5d_{3/2}\rangle=3.08$ and $\langle 6p_{3/2} || D || 5d_{3/2}\rangle=1.32$, we get 
from $R^{\lambda=514.53nm} = -11.57$ and $R^{\lambda=1111.68nm} = 0.409$ to $R^{\lambda=514.53nm} =-11.36$ and $R^{\lambda=1111.68nm} = 0.418$. By fixing 
$\langle 4f_{5/2} || D || 5d_{3/2}\rangle=4.00$, when we reshuffle other matrix 
elements, we get light-shift ratios close to the experimental values for
 $\langle 6p_{1/2} || D || 6s\rangle=3.35$, $\langle 6p_{3/2} || D || 6s\rangle=4.72$, $\langle 6p_{1/2} || D || 5d_{3/2}\rangle=3.08$ and $\langle 6p_{3/2} || D || 5d_{3/2}\rangle=1.34$. 
When we keep these matrix elements constant and vary the matrix element of $\langle 4f_{5/2} || D || 5d_{3/2}\rangle$, we get the
best light-shift ratios for 3.65 au and they correspond to $R^{\lambda=514.53nm} = -11.49$ and $R^{\lambda=1111.68nm} = 0.419$. For any other combinations, 
the light-shift ratios vary by large amount from the experimental results;
at least for one of the wavelengths. 

 Again, there are also measurements of the transition probabilities in Ba$^+$
available from three different experiments \cite{kastberg,reader,gallagher1}. 
We evaluate them using the experimental wavelengths (determined from the 
corresponding experimental excitation energies \cite{moore}) and E1 matrix
elements presented in Table \ref{tab1}. These results are given in Table
\ref{tab3}. We have also presented both the experimental and other theoretical
results in the same table. It can be clearly noticed that our results in 
all the cases match well with the experimental results with smaller error bars
where as some of the earlier works differ significantly. Differences
between all the theoretical works are discussed by Geetha et al. \cite{geetha}.
\begin{table}[h]
\caption{Transition strengths (au), wavelengths (\AA) and probabilities ($\times 10^6s^{-1}$) from P-states in Ba$^+$.}
\begin{ruledtabular}
\begin{center}
\begin{tabular}{lccccc}
Transition($f\rightarrow i$) & $S_{f\rightarrow i}^{E1}$ & $\lambda_{f\rightarrow i}^{\dagger}$ & $A_{f\rightarrow i}^{E1}$ & Others & Expt.\\
\hline \\
   &  & & & & \\
$6\text{P}_{1/2}\rightarrow 6\text{S}_{1/2}$ & 11.290  & 4935.5 & 95.131 & 93.68$^a$ & 95(9)$^c$ \\
 &   &  &  & 91.78$^b$ & 95.5(10)$^d$ \\
 &   &  &  &  & 95(7)$^e$ \\
$A_{6\text{P}_{3/2}\rightarrow 6\text{S}_{1/2}}$ & 22.373  & 4555.3 & 119.889 & 119.37$^a$ & 106(9)$^c$\\
 &   &  &  & 116.25$^b$ & 117(4)$^d$ \\
 &   &  &  &  & 118(8)$^e$ \\
$A_{6\text{P}_{1/2}\rightarrow 5\text{D}_{3/2}}$ & 9.672  & 6498.7 & 35.701 & 32.609$^a$ & 33.8(19)$^c$ \\
 &   &  &  & 33.42$^b$ & 33(8)$^d$ \\
 &   &  &  &  & 33(4)$^e$ \\
$A_{6\text{P}_{3/2}\rightarrow 5\text{D}_{3/2}}$ & 1.796  & 5855.3 & 4.531 & 4.255$^a$ & 4.69(29)$^c$\\
 &   &  &  & 3.342$^b$ & 4.8(5)$^d$ \\
 &   &  &  &  & 4.48(6)$^e$ \\
$A_{6\text{P}_{3/2}\rightarrow 5\text{D}_{5/2}}$ & 16.160  & 6143.4 & 35.305 & 34.93$^a$ & 37.7(24)$^c$ \\
 &   &  &  & 35.95$^b$ & 37(4)$^d$ \\
 &   &  &  &  & 37(4)$^e$ \\
\end{tabular}
\end{center}
\end{ruledtabular}
\label{tab3}
$^{\dagger} \lambda$s are determined from the experimental excitation energies \cite{moore}.\\ 
References: $^a$\cite{geetha}, $^b$\cite{johnson}, $^c$\cite{kastberg}, $^d$\cite{reader} and $^e$\cite{gallagher1}.
\end{table}

Now if we consider $\langle 6p_{1/2} || D || 6s \rangle$ as 3.35 au as we 
analyzed from the light-shift ratio studies, we get $A_{6\text{P}_{1/2}\rightarrow 6\text{S}_{1/2}}$ as $94.566 \times 10^6s^{-1}$ which is at the marginal 
lower side error limit of Reader et al experimental result \cite{reader}. 
Our $A_{6\text{P}_{1/2}\rightarrow 5\text{D}_{3/2}}$ result is also at the
marginal upper limit of Kastberg et al experimental result \cite{kastberg}.
If we consider $\langle 6p_{1/2} || D || 5d_{3/2} \rangle=3.08$ from the 
analysis of light-shift ratio studies then we get $A_{6\text{P}_{1/2}\rightarrow 5\text{D}_{3/2}}=35.015 \times 10^6s^{-1}$ which is within the experimental 
error bar.

By gathering all the informations from the analysis of roles of various matrix 
elements in the light-shift ratios and transition probabilities studies, we arrive
in the conclusion at this point that the magnitude of the following matrix 
elements will be
\begin{eqnarray}
\langle 6p_{1/2} || D || 6s_{1/2}\rangle&=&3.36(2) \nonumber \\
\langle 6p_{3/2} || D || 6s_{1/2}\rangle&=&4.73(3) \nonumber \\
\langle 6p_{1/2} || D || 5d_{3/2}\rangle&=&3.08(3) \nonumber \\
\langle 6p_{3/2} || D || 5d_{3/2}\rangle&=&1.34(2) \nonumber \\
\langle 4f_{5/2} || D || 5d_{3/2}\rangle&=&3.73(20). \nonumber
\end{eqnarray}
The above conclusions differ from the findings of the previous work by
Sherman et al. \cite{fortson4} who find
\begin{eqnarray}
\langle 6p_{1/2} || D || 5d_{3/2}\rangle&=&3.14(3) \nonumber \\
\langle 4f_{5/2} || D || 5d_{3/2}\rangle&=&4.36(36), \nonumber
\end{eqnarray}
where they had combined with Geetha et al \cite{geetha} results to estimate these
values. Further more, light-shift ratio measurements with a large number of 
wavelengths are required in order to estimate the above E1 matrix elements 
more accurately.

\begin{table}[h]
\caption{Net transition probabilities ($\times 10^6s^{-1}$) and lifetimes ($ns$) in Ba$^+$.}
\begin{ruledtabular}
\begin{center}
\begin{tabular}{lcccc}
State($=i$) & $A_{i}$ & $\tau_{i}$ & Others & Expt. \\  
\hline \\
&  & & & \\
$6\text{P}_{1/2}$ & 130.146 & 7.68(7) & 7.89$^a$ & 7.92(8)$^e$\\
&  & & 7.92$^b$ & 7.90(10)$^f$\\
&  & & 7.83$^c$ & \\
&  & & 7.99$^d$ & \\
$6\text{P}_{3/2}$ & 159.72 & 6.26(11) & 6.30$^a$& 6.32(10)$^f$\\
&  & & 6.31$^b$ & 6.312(16)$^g$\\
&  & & 6.27$^c$ & \\
&  & & 6.39$^d$ & \\
\end{tabular}
\end{center}
\end{ruledtabular}
\label{tab4}
References: $^a$\cite{dzuba}, $^b$\cite{geetha}, $^c$\cite{safronova1}, $^d$\cite{johnson}, $^e$\cite{kuske}, $^f$\cite{pinnington} and $^g$\cite{andrae}.
\end{table}
There are a number of measurements on the lifetimes of the 
$6p \ ^2\text{P}_{1/2}$ and $6p \ ^2\text{P}_{3/2}$ states are available in 
the literature \cite{kuske,pinnington,andrae}, but the associate error bars
of these measurements are very large. We evaluate the net transition 
probabilities of these states using the above matrix elements and experimental
wavelengths \cite{moore}. We give these results in Table \ref{tab4} and 
compare with the corresponding experimental results. Results obtained from
other works are also given in the same table. As it can be seen from this
table that the lifetime of the $6\text{P}_{1/2}$ state is more than 2\%
 difference 
from the earlier findings which needs to be verified by new experiments. 
Although, our calculation of the lifetime of the $6\text{P}_{3/2}$ state
is within the error bar of the experimental results and agree with other
findings, but the $A_{6\text{P}_{3/2}\rightarrow 5\text{D}_{5/2}}$ contributes
around 28\% in its evaluation. Since the accuracy of the E1 matrix element of 
$\langle 6p_{3/2} || D || 5d_{5/2}\rangle$ is not verified well as it has been 
done
for the transition matrix elements involved in the determination of lifetime
of the  $6\text{P}_{1/2}$ state, the lifetime of the $6\text{P}_{3/2}$ state 
may be little different than what we find, but it will be within the error
bar that we have given (which is large). The error bars given in our lifetime 
calculations come only from the error bars of the E1 matrix elements as 
given above.

\begin{table*}[h]
\caption{Scalar and tensor static ($\omega=0$) polarizabilities in Ba$^+$.}
\begin{ruledtabular}
\begin{center}
\begin{tabular}{lcccccccc}
Intermediate & $6\text{S}_{1/2}$ & $6\text{P}_{1/2}$ & \multicolumn{2}{c}{$6\text{P}_{3/2}$} & \multicolumn{2}{c}{$5\text{D}_{3/2}$} & \multicolumn{2}{c}{$5\text{D}_{5/2}$} \\
States ($K$)& Scalar &  Scalar  &  Scalar & Tensor &  Scalar & Tensor &  Scalar & Tensor \\
\hline \\
$6\text{S}_{1/2}$ &        & $-$40.763 & $-$37.280 & 37.280    &       &  \\
$7\text{S}_{1/2}$ &        &  19.714   & 25.889    & $-$25.889 &       &  \\
$8\text{S}_{1/2}$ &        &  0.844    & 0.954     & $-$0.954  &       &  \\
$6\text{P}_{1/2}$ & 40.763 &           &           &           & 22.992& $-$22.992 & &  \\
$7\text{P}_{1/2}$ &  0.015 &           &           &           & 0.064 & $-$0.064 & &  \\
$8\text{P}_{1/2}$ &  0.014 &           &           &           & 0.011 & $-$0.011 &  &  \\
$6\text{P}_{3/2}$ & 74.560 &           &           &           & 3.846 & 3.077 & 24.210 &$-$24.210  \\
$7\text{P}_{3/2}$ &  0.043 &           &           &           & 0.021 & 0.017 & 0.116 &$-$0.116  \\
$8\text{P}_{3/2}$ &  0.014 &           &           &           & 0.003 & 0.003 & 0.019 & $-$0.019 \\
$5\text{D}_{3/2}$ &        & $-$45.984 & $-$3.846  & $-$3.077  &       &  \\
$6\text{D}_{3/2}$ &        & 68.101    & 8.275     & 6.620     &       &  &  &  \\
$7\text{D}_{3/2}$ &        &  4.163    & 0.434     & 0.347     &       &  &  &  \\
$5\text{D}_{5/2}$ &        &           & $-$36.315 & 7.263     &       &  &  &  \\
$6\text{D}_{5/2}$ &        &           & 72.167    & $-$14.433 &       &  &  &  \\
$7\text{D}_{5/2}$ &        &           & 3.895     & $-$0.779  &       &  &  &  \\
$4\text{F}_{5/2}$ &        &           &           &           & 11.857 & $-$2.371 & 0.668 & 0.763 \\
$5\text{F}_{5/2}$ &        &           &           &           & 1.643  & $-$0.329 & 0.089 & 0.102 \\
$6\text{F}_{5/2}$ &        &           &           &           & 0.017  & $-$0.003 & 0.009  & 0.010 \\
$4\text{F}_{7/2}$ &        &           &           &           &       &  & 13.344 & $-$4.766 \\
$5\text{F}_{7/2}$ &        &           &           &           &       &  & 2.655 & $-$0.948 \\
$6\text{F}_{7/2}$ &        &           &           &           &       &  & 0.430 & $-$0.154 \\
Others  & 9.137 & 11.304 & 11.716 & $-$0.508 & 8.981 & $-$.019 & 9.000 & $-$.084 \\
\hline \\
Total  & 124.546(1.256) & 17.379(2.920) & 45.890(3.831) & 5.870(128) & 49.435(1.458) & $-$22.694(538) & 50.540(1.410) & $-$29.422(931) \\
Expt. \cite{snow} &  123.88(5)      &   &      &       &   &  &  &  \\
 \ \ \ \ \ \ \ \ \ \cite{gallagher2}  & 125.5(10)    &   &     &      &   &  &  &  \\
Theory \cite{safronova1}  & 124.15  &   &    &    &   &  &  &  \\
\ \ \ \ \ \ \ \ \ \ \cite{lim}  & 123.07  &   &    &    &   &  &  &  \\
\ \ \ \ \ \ \ \ \ \ \cite{miadokova}  & 126.2  &   &    &    &   &  &  &  \\
\ \ \ \ \ \ \ \ \ \ \cite{patil}  & 124.7  &   &    &    &   &  &  &  \\
\end{tabular}
\end{center}
\end{ruledtabular}
\label{tab5}
\end{table*}
Since the knowledge of the static dipole polarizabilities of different states 
in Ba$^+$ is important to estimate shifts in the energy levels when they are
subjected to high precision experiments like atomic clock, PNC and so on, we
have calculated these quantities for the first five low-lying states in the 
given system. Likewise vector polarizabilities, the scalar and tensor static
dynamic polarizabilities are evaluated using our E1 matrix elements presented 
in Table \ref{tab1} and their experimental excitation energies \cite{moore}.
We have considered the $6\text{S}_{1/2}$, $7\text{S}_{1/2}$, $8\text{S}_{1/2}$,
$5\text{D}_{3/2}$, $6\text{D}_{3/2}$, $7\text{D}_{3/2}$, $5\text{D}_{5/2}$,
$6\text{D}_{5/2}$, $7\text{D}_{5/2}$, $4\text{F}_{7/2}$, $5\text{F}_{7/2}$ and 
$6\text{F}_{7/2}$ states in these calculations 
along with the intermediate states considered for the above vector 
polarizabilities calculations. Again, we have calculated the core
correlation effects using the relativistic CCSD method employed for the 
closed-shell system in our earlier work \cite{bijaya6}. The core-valence and 
higher state contributions other than
the matrix elements reported in Table \ref{tab1} are evaluated using the 
MBPT(2) method. Our static dipole polarizability results of various states 
in Ba$^+$ are presented in Table \ref{tab5}. The core-correlation effects
for the scalar and tensor polarizabilities are 9.582 au and $-$0.372 au,
respectively. There are experimental results available only for the ground
state \cite{snow,gallagher2}. Again, a couple of theoretical calculations
available for the ground state \cite{safronova1,lim,miadokova,patil}. Except
Iskrenova-Tchoukova and Safronova work \cite{safronova1}, others have employed molecular codes
to determine them. Also, Iskrenova-Tchoukova and Safronova have used the E1
matrix elements from the linearized CCSD(T) method to evaluate the ground state
polarizabilities. In their work, they have considered only four important 
matrix elements to evaluate them and we have considered two more higher excited
states matrix elements in this calculation. Again, their core-correlation 
effects are considered using the lower order many-body methods and it is 
larger than
our finding. There are no other results for the excited states available
to compare with our results. The error bars given in our calculations are
from the errors given for the E1 matrix elements in Table \ref{tab1}.
The agreement between our ground state polarizability result with its 
experimental measurement further supports the accuracy of the E1 matrix 
elements used in this calculation.

\subsection{Ra$^+$}
\begin{table}[h]
\caption{Absolute magnitudes of the reduced dipole matrix elements in Ra$^{+}$. Estimated error bars from this work given inside the parenthesis.}
\begin{ruledtabular}
\begin{center}
\begin{tabular}{lccc}
Transition & Present & Ref. \cite{dzuba} & Ref. \cite{safronova2} \\
\hline \\
   &  &  &  \\
$7p_{1/2}\rightarrow 7s_{1/2}$ & 3.28(2) & 3.224 & 3.2545 \\
$8p_{1/2}\rightarrow 7s_{1/2}$ & 0.08(4) & 0.088 &     \\
$9p_{1/2}\rightarrow 7s_{1/2}$ & 0.09(3) & 0.116 &     \\
$7p_{3/2}\rightarrow 7s_{1/2}$ & 4.54(2) & 4.477 & 4.5106 \\
$8p_{3/2}\rightarrow 7s_{1/2}$ & 0.49(2) & 0.339 &     \\
$9p_{3/2}\rightarrow 7s_{1/2}$ & 0.30(2) & 0.095 &     \\
$7p_{1/2}\rightarrow 6d_{3/2}$ & 3.62(5) & 3.550 & 3.5659 \\
$8p_{1/2}\rightarrow 6d_{3/2}$ & 0.06(2) & 0.013 &     \\
$9p_{1/2}\rightarrow 6d_{3/2}$ & 0.02(1) & 0.013 &     \\
$7p_{3/2}\rightarrow 6d_{3/2}$ & 1.54(2) & 1.504 & 1.5117 \\
$8p_{3/2}\rightarrow 6d_{3/2}$ & 0.15(2) & 0.127 &     \\
$9p_{3/2}\rightarrow 6d_{3/2}$ & 0.07(2) & 0.057 &     \\
$5f_{5/2}\rightarrow 6d_{3/2}$ & 4.67(2) &    & 4.4491 \\
$6f_{5/2}\rightarrow 6d_{3/2}$ & 0.86(4) &    &     \\
$7f_{5/2}\rightarrow 6d_{3/2}$ & 0.48(11) &    &     \\
$7p_{3/2}\rightarrow 6d_{5/2}$ & 4.83(8) & 4.816 & 4.8232 \\
\end{tabular}
\end{center}
\end{ruledtabular}
\label{tab6}
\end{table}

We have also followed the same procedure as above to determine the light-shift 
ratios and lifetimes of the first excited P-states in Ra$^+$. We present
the various E1 matrix elements used in these calculations in Table \ref{tab6}.
The error bars of different matrix elements are estimated using the same 
procedure followed as in Ba$^+$. There are also other
calculations available in this system \cite{dzuba,safronova2}. Dzuba et al.
\cite{dzuba} have used the same method as in Ba$^+$ to calculate these 
matrix elements. The procedure followed by Safronova et al. \cite{safronova2}
is based on the linearized CCSD(T) method like Iskrenova-Tchoukova and Safronova
work in Ba$^+$ \cite{safronova1}. Most of our results match with these 
calculations, but they still differ at the second decimal places as in
Ba$^+$.

Since Ra$^+$ has been proposed for the atomic clock \cite{bijaya2} and
PNC experiments \cite{wansbeek1}, knowledge of the accuracies of the above
E1 matrix elements are essential. 
We propose also similar light-shift ratio measurements at different wavelengths
for the same purpose. As a preliminary study, we consider the same off-resonant
wavelengths as have been considered in the Ba$^+$ experiments 
\cite{fortson3,fortson4} and have carried out light-shift ratio calculations. 
In Table \ref{tab7}, we present the dynamic vector polarizabilities for the  
$7\text{S}_{1/2}$ and $6\text{D}_{3/2}$ states at $\lambda=514.53$nm and $\lambda=1111.68$nm. We use our E1 matrix elements presented in Table \ref{tab6} with 
the experimental energies \cite{moore} to evaluate them. In this calculation,
we have considered the $7\text{P}_{1/2}$, $8\text{P}_{1/2}$, $9\text{P}_{1/2}$, 
$7\text{P}_{3/2}$, $8\text{P}_{3/2}$, $9\text{P}_{3/2}$, $5\text{F}_{5/2}$,
$6\text{F}_{5/2}$ and $7\text{F}_{5/2}$ intermediate states.
\begin{table*}[h]
\caption{Dynamic vector polarizabilities (in au) at two different wavelengths in Ra$^{+}$.}
\begin{ruledtabular}
\begin{center}
\begin{tabular}{lcccc}
Intermediate & \multicolumn{2}{c}{$7s_{1/2}$} & \multicolumn{2}{c}{$6d_{3/2}$} \\
States ($K$)  &  $\lambda=514.53$nm  &  $\lambda=1111.68$nm &  $\lambda=514.53$nm  &  $\lambda=1111.68$nm \\
\hline \\
   &  &  &  \\
$7p_{1/2}$ & $-$391.483 & $-$37.764 &  31.921 & $-$869.894 \\
$8p_{1/2}$ & $-$0.008 & $-$0.003 & $-$0.002  & $-8\times10^{-4}$   \\
$9p_{1/2}$ & $-$0.007 & $-$0.003 & $-1\times10^{-4}$ & $-5\times10^{-5}$   \\
$7p_{3/2}$ & 94.788 & 22.384 & 3.784 & $-$2.632 \\
$8p_{3/2}$ & 0.144 & 0.059 & $-$0.005 & $-$0.002   \\
$9p_{3/2}$ & 0.035 & 0.015 & $-6\times10^{-4}$ & $-3\times10^{-4}$   \\
$5f_{5/2}$ &  &    & 9.452 & 3.361 \\
$6f_{5/2}$ &  &    & 0.149 &  0.061   \\
$7f_{5/2}$ &  &    & 0.035 & 0.015   \\
Others       & 0.323 & 0.145  & 0.795 & 0.345 \\
\hline \\
Total & $-$296.207(4.909) & $-$15.167(255) & 46.128(679) & $-$868.749(24.027) \\
\end{tabular}
\end{center}
\end{ruledtabular}
\label{tab7}
\end{table*}

As seen in Table \ref{tab7}, the trend of the contributions from various 
intermediate states is same as in Ba$^+$. The core-correlations are little
larger compared to Ba$^+$ and they are around 0.346 au and 0.159 au at 
$\lambda=514.53$nm and $\lambda=1111.68$nm, respectively. 
We obtain around $R^{\lambda=514.53nm} = -6.42(7)$ and $R^{\lambda=1111.68nm} = 0.017(1)$. The given error bars again 
come from our estimated accuracies of the E1 matrix elements. Since there are 
no experimental data available in Ra$^+$ to check the accuracy of the E1 matrix 
elements or light-shift ratios, our results will be useful for the future
light-shift experiments in this system.  
\begin{table}[h]
\caption{Transition strengths (au), wavelengths (\AA) and probabilities ($\times 10^6s^{-1}$) from P-states in Ra$^+$.}
\begin{ruledtabular}
\begin{center}
\begin{tabular}{lccc}
Transition($f\rightarrow i$) & $S_{f\rightarrow i}^{E1}$ & $\lambda_{f\rightarrow i}$ & $A_{f\rightarrow i}^{E1}$  \\
\hline \\
  &   & & \\
$A_{7\text{P}_{1/2}\rightarrow 7\text{S}_{1/2}}$ & 11.086 & 4683.6 & 106.083 \\
$A_{7\text{P}_{3/2}\rightarrow 7\text{S}_{1/2}}$ & 20.612 & 3815.5 & 187.960 \\
$A_{7\text{P}_{1/2}\rightarrow 6\text{D}_{3/2}}$ & 13.104 & 10791.2 & 10.564 \\
$A_{7\text{P}_{3/2}\rightarrow 6\text{D}_{3/2}}$ & 2.372  & 7080.0 & 3.385  \\
$A_{7\text{P}_{3/2}\rightarrow 6\text{D}_{5/2}}$ & 23.329 & 8022.0  & 22.890 \\
\end{tabular}
\end{center}
\end{ruledtabular}
\label{tab8}
\end{table}

We have already reported polarizabilities of the $7\text{S}_{1/2}$, $6\text{D}_{3/2}$ and $6\text{D}_{5/2}$ states in our earlier work \cite{bijaya2}. However,
the knowledge of the transition probabilities and lifetimes of the first 
P-states in the considered states are not known yet. We use our E1 matrix 
elements reported in Table \ref{tab6} and experimental energies \cite{moore} 
to evaluate them. In Table \ref{tab8}, we present the transition probabilities 
of various channels from the $7\text{P}_{1/2}$ and $7\text{P}_{3/2}$ states.
As seen from this table, the relative transition probability of the $7\text{P}_{3/2}\rightarrow 7\text{S}_{1/2}$ transition is much larger than the $7\text{P}_{1/2}\rightarrow 7\text{S}_{1/2}$ transition in the present system compared to Ba$^+$. Since the wavelength of the $7\text{P}_{1/2}\rightarrow 6\text{D}_{3/2}$ transition is one order larger here, 
the corresponding transition probability is comparatively smaller. Hence, the
lifetime of the $7\text{P}_{1/2}$ state becomes little larger than the $6\text{P}_{1/2}$ state in Ba$^+$. Again, the lifetime of the $7\text{P}_{3/2}$ state
becomes smaller than the $6\text{P}_{3/2}$ state in Ba$^+$ due to the fact
that the $7\text{P}_{3/2}\rightarrow 7\text{S}_{1/2}$ transition probability
is much larger. We present the net transition probabilities and lifetimes
of the $7\text{P}_{1/2}$ and $7\text{P}_{3/2}$ states in Table \ref{tab9}.
In Ra$^+$, $A_{7\text{P}_{3/2}\rightarrow 6\text{D}_{5/2}}$ contributes
only by 12\% to the final lifetime determination of the $7\text{P}_{3/2}$ state
 in contrast to 28\% in Ba$^+$.

\begin{table}[h]
\caption{Net transition probabilities ($\times 10^6s^{-1}$) and lifetimes ($ns$) in Ra$^+$.}
\begin{ruledtabular}
\begin{center}
\begin{tabular}{lcc}
State($=i$) & $A_{i}$ & $\tau_{i}$  \\  
\hline \\
  &  & \\
$7\text{P}_{1/2}$ & 116.647 & 8.57(12) \\
$7\text{P}_{3/2}$ & 214.235 & 4.67(5)  \\
\end{tabular}
\end{center}
\end{ruledtabular}
\label{tab9}
\end{table}

\section{Conclusion}
We have employed the relativistic coupled-cluster theory to calculate the
electric dipole matrix
elements in the singly ionized barium and radium. These elements are used
to determine the light-shift ratios and transition probabilities in the singly
ionized barium and comparing with the corresponding experimental data, we
have estimated the accuracy of various low-lying matrix elements in this
system. Further, we have evaluated the scalar and tensor static dipole 
polarizabilities and lifetimes in the same system. We have also calculated the
light-shift ratios, transition probabilities and lifetimes using our electric
dipole matrix elements in the singly ionized radium. These data will be helpful
in the future experiments in the considered system.

\section{Acknowledgment}
This work is supported by NWO under VENI fellowship scheme with project number 
680-47-128 and put of the stichting FOM physics program 48 (Tri{$\mu$}p).
We thank Norval Fortson, Jeff Sherman, Bhanu Pratap Das and Dimtry Budker for useful discussions.

\end{document}